\documentclass[aps,twocolumn,superscriptaddress]{revtex4-1}
\usepackage{graphics}
\usepackage{epsfig}
\usepackage[usenames]{color}
\usepackage{verbatim}
\usepackage{amsmath}
\usepackage{hyperref}
\IfFileExists{srcltx.sty}{\usepackage[active]{srcltx}}

\begin{document}

\title{Phenomenology of Supersymmetric Models with a Symmetry-Breaking Seesaw Mechanism}

\author{Lauren Pearce}
\affiliation{Department of Physics and Astronomy, University of California, Los Angeles, CA 90095-1547, USA}

\author{Alexander Kusenko}
\affiliation{Department of Physics and Astronomy, University of California, Los Angeles, CA 90095-1547, USA}
\affiliation{Kavli IPMU (WPI), University of Tokyo, Kashiwa, Chiba 277-8568, Japan}

\author{R. D. Peccei}
\affiliation{Department of Physics and Astronomy, University of California, Los Angeles, CA 90095-1547, USA}

\begin{abstract}
We explore phenomenological implications of the minimal supersymmetric standard model (MSSM) with a strong supersymmetry breaking trilinear term.  Supersymmetry breaking can trigger electroweak symmetry breaking via a symmetry-breaking seesaw mechanism, which can lead to a low-energy theory with multiple composite Higgs bosons.  In this model, the electroweak phase transition can be first-order for some generic values of parameters.  Furthermore, there are additional sources of CP violation in the Higgs sector. This opens the possibility of electroweak baryogenesis in the strongly coupled MSSM.   The extended Higgs dynamics can be discovered at Large Hadron Collider or at a future linear collider.

\end{abstract}

\maketitle

\section{Introduction}

The minimal supersymmetric standard model (MSSM) provides an appealing framework for physics beyond the Standard Model.  However, the lack of direct experimental evidence for superpartners, as well as the recent discovery of a Higgs boson with a mass that is heavier than one would naively expect in the MSSM, challenges the viability of low-energy supersymmetry in its simplest realizations. At the same time, there exists a possibility for a strongly coupled form of MSSM, which is associated with a large value of the trilinear supersymmetry breaking coupling~\cite{Kusenko:1998yj,Giudice:1998dj,Cornwall:2012ea}.  In this strongly coupled regime, the squarks form bound states via the exchange of Higgs bosons, creating some additional composite states, which can have some non-zero vacuum expectation values (VEVs).  

This class of models has several intriguing features.  First, the particle content of the low-energy effective theory is different from what one expects in the MSSM: in particular, the model predicts more Higgs bosons and fewer superpartners to be observed at the electroweak scale.   Second, the bound state quartic coupling is not related to the gauge coupling, which relaxes the upper bound on the mass of the lightest Higgs boson~\cite{Cornwall:2012ea}.  Third, the model contains gauge singlet states, whose presence allows for the electroweak transition to be first order; this model also has additional sources of CP violation.  The latter creates sufficient conditions for electroweak baryogenesis, as we discuss below. 

In this paper we will explore the ramifications of strongly coupled MSSM for electroweak baryogenesis, as well as collider phenomenology.

\section{Symmetry Breaking Seesaw}
\label{sec:Review_Seesaw}
\begin{figure}
\includegraphics[scale=.7]{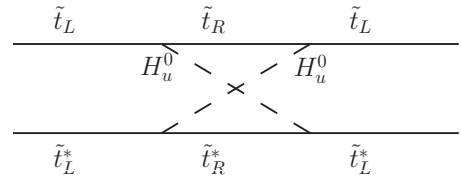}
\caption{The kernel of the bound state Higgs doublet.}
\label{fig:Doublet}
\end{figure}

Let us briefly review the strongly coupled phase of the MSSM~\cite{Kusenko:1998yj,Giudice:1998dj,Cornwall:2012ea}.  Supersymmetry breaking introduces the following trilinear terms in the Lagrangian:
\begin{equation}
A_t H_u \tilde{t}_L^* \tilde{t}_R + h.c.,
\end{equation}
and a similar term in the down sector.  If the coupling $A_t$ is sufficiently large, the stop squarks present in the theory can form bound states.  One of these bound states is an $SU_L(2)$ doublet with the same quantum numbers as the fundamental Higgs boson $H_u$.  The kernel of this state is shown in Fig. \ref{fig:Doublet}.  Although it is a crossed kernel, as opposed to the more familiar ladder diagram, the mass of the bound state can be found numerically as a function of $A_t$, and sufficiently large couplings can cause the mass squared of the bound state, $m_{BS}^2$, to become negative~\cite{Cornwall:2012ea}.  If this happens, electroweak symmetry breaking occurs.  It is an interesting feature of the model that electroweak symmetry breaking is triggered by supersymmetry breaking in a way that is fundamentally different from the weakly coupled MSSM. 

The discussion in Refs.~\cite{Kusenko:1998yj,Giudice:1998dj} and, in part, in Ref.~\cite{Cornwall:2012ea}, considered such large values of $A_t$ for which $m_{BS}^2<0$.  However, as was pointed out in Ref.~\cite{Cornwall:2012ea}, this is a sufficient, but not necessary condition.  It is possible that supersymmetry breaking would trigger electroweak symmetry breaking even for some much lower values of $A_t$.  

The bound-state composite Higgs doublet generically mixes with the fundamental Higgs doublet, due to the same $A_t H_u \tilde{t}_L^* \tilde{t}_R$ coupling. Thus the mass matrix, in the basis of $H_u$ and the bound state, has the form 
\begin{equation}
M^2 = \begin{pmatrix} m_h^2 & \alpha A_t^2 \\ \alpha^* A_t^{* \, 2} & m_{BS}^2 \end{pmatrix}, 
\end{equation}
where $\alpha$ is a numerical constant determined by strong dynamics, $m_h$ is the mass term of the fundamental Higgs doublet, and $m_{BS}$ is the mass of the bound state doublet.  A negative eigenvalue in this mass matrix signals that electroweak symmetry is broken, which can happen when both diagonal elements are positive. One of the eigenvalues is negative when
\begin{equation}
m_h^2 m_{BS}^2 - |\alpha|^2 |A_t|^4 < 0,
\end{equation}
which is possible for any positive values of $m_{BS}$ and $A_t$, provided that $m_h$ is a small enough (positive) number. The possibility of breaking electroweak symmetry for a relatively small $A_t$, smaller than the value required to drive the bound state mass to zero, opens a broad range of possibilities in the MSSM, which were not considered in Refs.~\cite{Kusenko:1998yj,Giudice:1998dj}.  This type of symmetry breaking, which relies on the mixing and the mass matrix similar to the seesaw neutrino mass matrix, was dubbed a {\em symmetry-breaking seesaw mechanism}~\cite{Cornwall:2012ea}.   

An additional benefit of the symmetry-breaking seesaw mechanism is that it automatically preserves $SU_C(3)$.   The two squarks in the bound state carry color, and the bound states of the form shown in Fig.~\ref{fig:Doublet} include an $SU_C(3)$ octet along with the color singlet.  If the octet acquires a non-zero VEV, it would break $SU_C(3)$ making the model unacceptable.  However, to acquire a VEV, the bound state must mix with the fundamental Higgs boson, which is only possible for an $SU_C(3)$ singlet state.   The quantum numbers of the fundamental Higgs boson of the MSSM dictate the pattern of symmetry breaking by selecting the only state that is phenomenologically acceptable, among the numerous possibilities.

\section{Description of Bound States}
\label{sec:States}

\begin{figure}
\includegraphics[scale=.7]{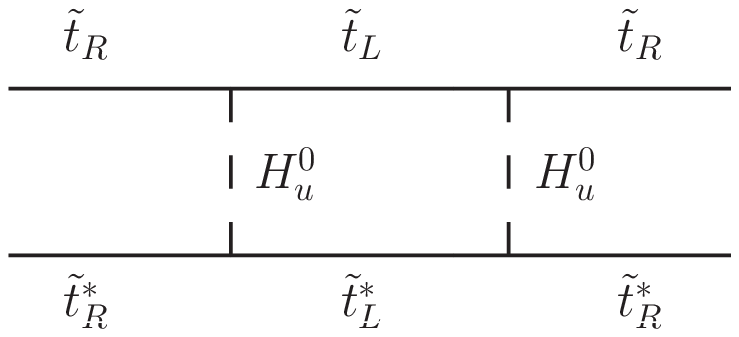}
\includegraphics[scale=.7]{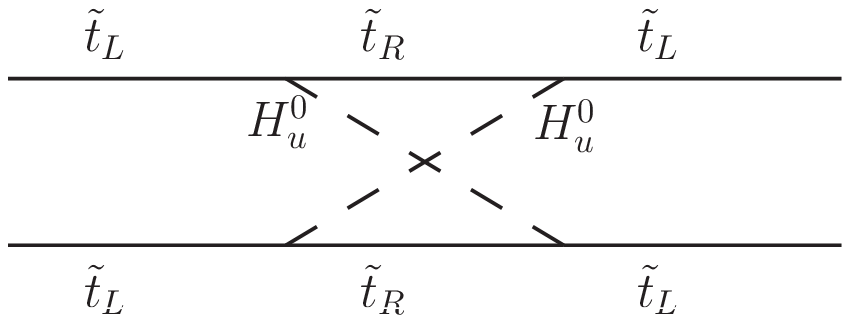}
\caption{Other kernels for squark bound states which produce $SU_L(2)$ singlets and triplets.}
\label{fig:Kernels}
\end{figure}

In Ref.~\cite{Cornwall:2012ea}, the primary focus was the possibility of spontaneous symmetry breaking; in this current work, we are interested in phenomenology.  Let us begin with a discussion of the bound states present in this model.  In addition to the $SU_L(2)$ doublet mentioned above, there are $SU_L(2)$ singlets and $SU_L(2)$ triplets; an example of the relevant kernels are shown in Fig. \ref{fig:Kernels}.  The vertices in these bound states are all proportional $|A_t|$; because Yukawa interactions are attractive in all channels, all of these states exist if the doublet bound state mentioned above exists.  We will assume that only $A_t$ (and possibly $A_b$) are large enough to produce bound states; these bound states appear in the up and down Higgs sectors respectively.  We have summarized the possible bound states, along with their quantum numbers, in Table \ref{tb:States}.

\begin{table}
\begin{ruledtabular}
\begin{small}
\begin{tabular}{|c|c|c|c|c|}
\hline & Squarks &$SU_L(2)$ Charge&$SU_C(3)$ Charge& $U_Y(1)$ Charge\\ 
\hline 1 & $\tilde{t}_L \tilde{t}_L$ & $2 \otimes 2 = 3 \oplus 1$ & $3 \otimes 3 = 6 \oplus \bar{3} $ & $1\slash 3$ \\
\hline 2 & $\tilde{t}_L^* \tilde{t}_L^*$ & $\bar{2} \otimes \bar{2} = \bar{3} \oplus 1$ & $\bar{3} \otimes \bar{3} = \bar{6} \oplus 3 $ & $-1\slash 3$ \\
\hline 3 & $\tilde{t}_R \tilde{t}_R$ & $1 \otimes 1 = 1$ & $3 \otimes 3 = 6 \oplus \bar{3} $ & $4 \slash 3$ \\
\hline 4 & $\tilde{t}_R^* \tilde{t}_R^*$ & $1 \otimes 1 = 1$ & $\bar{3} \otimes \bar{3} = \bar{6} \oplus 3 $ & $-4 \slash 3$ \\
\hline 5 & $\tilde{t}_L \tilde{t}_R^*$ & $2 \otimes 1 = 2$ & $3 \otimes \bar{3} = 1 \oplus 8$ & $-1 \slash 2$\\
\hline 6 & $\tilde{t}_R \tilde{t}_L^*$ & $\bar{2} \otimes 1 = \bar{2}$ & $3 \otimes \bar{3} = 1 \oplus 8$ & $1 \slash 2$ \\
\hline 7 & $\tilde{t}_L \tilde{t}_L^*$ & $2 \otimes \bar{2} = 3 \oplus 1 $ & $3 \otimes \bar{3} = 1 \oplus 8$ & 0 \\
\hline 8 & $\tilde{t}_R \tilde{t}_R^*$ & $1 \otimes 1 = 1 $ & $3 \otimes \bar{3} = 1 \oplus 8$ & 0 \\
\hline 
\end{tabular} 
\end{small}
\end{ruledtabular}
\caption{Quantum numbers of bound states; note that in $SU(2)$, the antifundamental representation $\bar{2}$ is identical to the fundamental representation $2$.  Line numbers are provided for ease of reference in the text.} 
\label{tb:States}
\end{table}

Let us summarize the states present in the model and their salient features.   We observe that rows 1 and 2 in Table~\ref{tb:States} are hermitian conjugates of each other; consequently, these rows may be combined to form complex representations.  Rows 3 and 4 may also be similarly combined.  Thus, the first two rows in the table describe a complex $SU_L(2)$ triplet and a complex $SU_L(2)$ singlet, while the next two rows describe another complex $SU_L(2)$ singlet.  All of these states carry color charge, and they all also have fractional electric charge.  The implications of these states will be discussed further in Section \ref{sec:collider} on collider phenomenology.

Similarly, the next two rows (5 and 6) are also hermitian conjugates of each other, which may be combined into a complex $SU_L(2)$ doublet.  This is the doublet discussed in Section \ref{sec:Review_Seesaw} above.  As mentioned, there are additionally 8 colored states which do not generally acquire vacuum expectation values.  The last two rows (7 and 8) describe two real singlets and one real triplet; these come in both colored and colorless versions.  As we will show in Section \ref{sec:EW_Phase_Transition}, the electroweak phase transition is generally first order due to these singlets.

Thus, we see that our model of strongly interacting supersymmetry has a rather extended Higgs sector.  With so many degrees of freedom, it is important to ensure that no colored states acquire a vacuum expectation value.  We note that there are no terms which involve only a single colored field, because the Lagrangian must be invariant under $SU_C(3)$.  Thus, even if all uncolored states acquire non-zero vacuum expectation values, there is no term linear in a colored field; such a term would necessarily induce spontaneous symmetry breaking of $SU_C(3)$.  These colored fields generally acquire corrections to their mass values during electroweak symmetry breaking; we will assume that their masses  sufficiently large that these corrections do not drive any of the mass-squared values negative.

The full model, including both up and down sectors, is rather complicated.  Therefore, we will make the simplifying assumption that the sectors are relatively decoupled, and we will only consider the up sector.  It is also possible that the bound states form only in the up sector.   We expect our results to hold in the more general model.  Next we will proceed to discuss the phenomenology of our model.  First, we will discuss flavor-changing-neutral-currents; this is a concern in any model in which extra Higgs doublets are present.  Second, we will consider the properties of the electroweak phase transition in this model.  Then we will discuss CP violation and baryogenesis, and, finally, we will make some remarks regarding collider phenomenology.

\section{Flavor Changing Neutral Currents}
\label{sec:FCNC}

Any model which introduces additional Higgs doublets must address the issue of flavor-changing-neutral-currents (FCNCs), which are highly constrained experimentally and generically large when additional $SU_L(2)$ doublets are introduced.  One well-known method of suppressing FCNCs is to have the doublets couple to different types of quarks; for example, in the MSSM the Higgs doublet $H_u$ only couples to up-type quarks and the doublet $H_d$ only couples to down-like quarks.  This suppresses FCNCs provided that the mixing between the two doublets (after supersymmetry is spontaneously broken) is not too large; this assumption that the sectors are relatively decoupled is made both by the MSSM and our model.

However, even if the up-sector and down-sector are decoupled, our model potentially has large FCNCs because we have two doublets within each sector; in particular, both the fundamental doublet $H_u$ and the bound state doublet $\Phi_u$ couple to up-type quarks.  Therefore, we consider a second way of suppressing FCNCs: if diagonalizing the quark matrix with respect to interactions with one of the doublets also approximately diagonalizes the quark matrix with respect to interactions with the other doublet, then FCNCs are small, because they are proportional to the off-diagonal elements.  Equivalently, FCNCs are  suppressed if the Yukawa couplings between the quarks and the first doublet are be approximately proportional to the Yukawa couplings between the quarks and the second doublet.  This approach is typically disfavored because it frequently requires fine-tuning, but we will demonstrate that this condition is naturally satisfied by our model.

We recall that the bound state doublet is comprised of up squarks exchanging $H_u$ bosons.  We note that there is no tree-level coupling between up-squarks and quarks; however, there is a tree-level coupling between $H_u$ and the quark: the Yukawa coupling $y_q$.  Therefore, to lowest order, an up-type quark sees only the $H_u$ contribution in the bound state, and so the coupling between the quark and the bound state is $y_q^\prime = \beta y_q$.

\begin{figure}
\includegraphics[scale=.7]{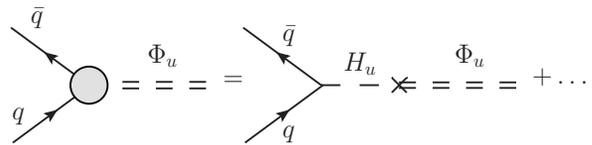}
\caption{The lowest order diagram for the Yukawa coupling between quarks and the bound state Higgs doublet.}
\label{fig:BS_quark_yukawa}
\end{figure}

The above argument is shown diagrammatically in Fig. \ref{fig:BS_quark_yukawa}; the lowest order contribution to the Yukawa coupling between a quark and the bound state Higgs doublet comes through the exchange of a true Higgs boson $H_u$.  Therefore, it is proportional to the Yukawa coupling $y_t$, and the proportionality constant $\beta$ describes the mixing between $H_u$ and the bound state $\Phi_u$.  This mixing $\beta$ is clearly independent of the quark involved on the right hand side of the diagram.  Therefore, the Yukawa couplings satisfy $y_q^\prime = \beta  y_q$, and hence FCNCs are indeed suppressed naturally, without fine-tunings.

\begin{figure}
\includegraphics[scale=.7]{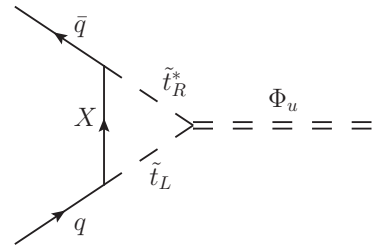}
\caption{The next order diagram for the Yukawa coupling between quarks and the bound state Higgs doublet.  The possibilities for particle X depends on the quarks involved, as discussed in the text.}
\label{fig:BS_quark_yukawa_2}
\end{figure}

We may be concerned about corrections from higher order diagrams; particularly in regards to the up-quark Yukawa coupling, which is quite small.  The next order corrections will come from diagrams like those shown in Fig. ~\ref{fig:BS_quark_yukawa_2}, in which particle X is a gaugino.  However, if the incoming quarks are up quarks, then the gaugino must convert an up quark into a stop squark, and the only gauginos which can change flavor is winos.  However, these are forbidden; winos can change up quarks only into down, strange, or bottom squarks.  Thus, there are no contributions from diagrams of this form to the up quark and charm quark Yukawa couplings.  Such diagrams do contribute to the top quark Yukawa coupling; for example, in this case the exchanged gaugino could be a gluino, Zino, photino, or Higgsino.  However, these are all smaller than the  first order contribution discussed above.

For future reference, let us relate $y_q$ and $y_q^\prime$ to the Yukawa couplings between the quark and the mass eigenstates; this will be relevant in our discussion of baryogenesis in Section \ref{sec:baryogenesis} below.  Let us assume that the mass eigenstates are related to $H_u$ and $\Phi_u$ by
\begin{align}
\Psi_1 &= \cos(\theta) H_u + \sin(\theta) \Phi_u \nonumber \\
\Psi_2 &= - \sin(\theta) H_u + \cos(\theta) \Phi_u. 
\end{align}
Then the relevant Yukawa couplings are given by
\begin{align}
y_{1q} &= \cos(\theta) y_q + \sin(\theta)y_q^\prime  \nonumber \\
&= \left( \cos(\theta) + \beta \sin(\theta) \right) y_q \nonumber \\
y_{2q} &= - \sin(\theta) y_q + \cos(\theta) y_q^\prime \nonumber \\
&= \left( - \sin(\theta) + \beta \cos(\theta) \right) y_q.
\end{align}
In particular we note that
\begin{equation}
y_{2q} = \dfrac{- \sin(\theta) + \beta \cos(\theta)}{\cos(\theta) + \beta \sin(\theta)} y_{1q},
\label{eq:yukawa}
\end{equation}
and we expect $\beta \sim \sin(\theta)$ due to its close relation to the mixing.

\section{Electroweak Phase Transition}
\label{sec:EW_Phase_Transition}

Let us briefly discuss the evolution of this model with temperature.  At sufficiently high temperatures, the model behaves as the standard (weakly interacting) MSSM.  At lower temperatures, the model undergoes a phase transition to a strongly interacting phase, and electroweak symmetry breaking takes place.  One can make an analogy with QCD, which is described by quarks and gluons at high temperature, but at some lower temperatures baryons and mesons become the appropriate degrees of freedom.  Likewise, in our model one should use the fundamental degrees of freedom of MSSM for temperatures well above a TeV, but one should consider bound states as new degrees of freedom at low temperatures.

In the strongly coupled phase, the model should be described by an effective Lagrangian written in terms of low-energy degrees of freedom, which include the bound states.  Ideally, one would like to calculate all the parameters in terms of the parameters of MSSM, which is the ultraviolet completion of the theory.  However, because the theory is strongly coupled, it is not feasible to calculate these parameters explicitly.  A calculation on the lattice may be possible~\cite{Hernandez:2001mi}, but no detailed results are available at present.  In the absence of such a calculation, one can only parametrize the low-energy couplings using generic values consistent with symmetries. (This is analogous to the approach that one took to strong interactions before QCD was discovered and understood.) Obviously, this approach is limited in what can be predicted. However, since the values of ``fundamental'' MSSM parameters in the high-energy Lagrangian are unknown and are not strongly constrained, the low-energy effective 
approach appears to be well justified.

We will now discuss the electroweak phase transition, which can be generically first-order in this model.  We note that the colorless $SU_L(2)$ gauge singlets are not associated with any symmetry breaking; therefore, they may acquire vacuum expectation values in the strongly coupled phase.  Such vacuum expectation values have no physical meaning and may be removed with a field redefinition that makes the tadpole diagrams vanish order-by order in perturbation theory.  We will assume that this has been done in writing the effective potential.  We have already shown that the colored fields do not acquire nonzero VEVs, and we will neglect them for the remainder of this section.  The effective potential for the colorless Higgs fields is written in Appendix \ref{sec:potential}; the cubic terms given in Equation \eqref{eq:Cubic_Terms} are particularly important for the first-order phase transition.  The notation used in the Appendix and this section is as follows: the complex $SU_L(2)$ doublet mass eigenstates are 
$\Psi_1$ and $\Psi_2$, where $\Psi_1$ has a negative mass-squared eigenvalue due to the seesaw symmetry breaking mechanism, the real $SU_L(2)$ singlet mass eigenstates are $S_1$ and $S_2$, and the real $SU_L(2)$ triplet field is $V$.

Due to the negative mass-squared of $\Psi_1$, the origin of the potential is not a local minimum and at least the neutral component of doublet $\Psi_1$ acquires a nonzero vacuum expectation value.  The terms $A_{S1} S_1 \Psi_1^\dagger \Psi_1$ and $\tilde{A}_{S1} S_2 \Psi_1^\dagger \Psi_1$ produce terms linear in $S_1$ and $S_2$ respectively, and so consequently $\left< S_1 \right> = 0$ nor $\left< S_2 \right> =0 $ can be a local minimum.  Hence, once $\Psi_1$ acquires a non-zero vacuum expectation value, the singlets also acquire a nonzero vacuum expectation value.  

When both $\Psi_1$ and the singlets have acquired non-zero VEVs, the neutral component of the other doublet, $\Psi_2$, must also acquire a nonzero vacuum expectation value due to terms such as $S_1 \Psi_1^\dagger \Psi_2$; the charged component does not acquire a nonzero vacuum expectation value.  Next let us consider the triplet; we parametrize it as $V^a = (V_1, V_2, V_3)$.  The cubic terms in Equation \eqref{eq:Cubic_Terms} include
\begin{align}
\Psi_1^\dagger \sigma^a V^a \Psi_1 &= \Psi_1^\dagger \begin{pmatrix} V_3 & V_1 - i V_2 \\ V_1 + i V_2 & V_3 \end{pmatrix} \Psi_1 \\
&= \Psi_1^\dagger \begin{pmatrix} V^0 \slash 2 & - V^+ \slash \sqrt{2} \\ - V^- \slash \sqrt{2} & -V^0 \slash 2 \end{pmatrix} \Psi_1,
\label{eq:triplet_matrix}
\end{align}
where we have identified the charge states.  When the neutral component of $\Psi_1$ acquires a nonzero vacuum expectation value, the above equation produces a term linear in $V^0$; consequently, this field also acquires a vacuum expectation value.  The consequences of this for the $\rho_0$ parameter and neutrino masses is discussed in Section \ref{sec:Triplet_Problems}.

The presence of gauge singlet Higgs states and the existence of tree-level cubic couplings generically make the phase transition strongly first-order. This is in contrast with the Standard Model and MSSM, in which the cubic terms are forbidden by the $SU_L(2)$ symmetry.  In the Standard Model, the transition is not first-order for the allowed range of the Higgs masses (more specifically, for the Higgs mass above 45 GeV).  In the MSSM, the transition is weakly first order, and only for such parameters for which the two-loop corrections generate a sufficient barrier in the potential~\cite{Carrington:1991hz,Dine:1992vs,Dine:1992wr,Giudice:1992hh}.  In our case, finite temperature corrections, calculated in the same manner as in the Standard Model \cite{Dolan:1973qd}, \cite{Weinberg:1974hy}, produce terms proportional to $T^2 M^2$, where $M^2$ are the mass eigenvalues of the shifted fields, as functions of the vacuum expectation values.  When the fields are shifted, the cubic terms produce terms in the mass 
eigenvalues linear in the vacuum expectation values.  Such linear terms produce a barrier which results in a strongly first order phase transition.  This is identical to that manner in which singlets produce a barrier in the Next-to-Minimal Supersymmetric Model (NMSSM)~\cite{Pietroni:1992in,Davies:1996qn,Huber:2000ih-fixed}.  

The potential written in Appendix \ref{sec:potential} has many parameters describing the cubic and quartic terms; this makes a comprehensive study of the temperature evolution of the potential impractical.  However, given the large number of parameters we expect there to be relatively large region of parameter space in which the phase transition occurs at a temperature of the order of the electroweak scale and $v_u = \sqrt{ |\left< \Psi_1 \right>|^2 + |\left< \Psi_2 \right>|^2} \lesssim 246$~GeV, with the difference to be made up in the down sector.  

\section{Baryogenesis}
\label{sec:baryogenesis}

The generic possibility of a strongly first-order phase transition reopens the possibility of baryogenesis at the electroweak scale~\cite{Kuzmin:1985mm,Rubakov:1996vz}.  In addition to the first-order phase transition, one needs CP violation for a successful baryogenesis.  The  full potential written in Appendix \ref{sec:potential} has 8 complex parameters; two of these may be eliminated by rotating the complex doublets $\Psi_1$ and $\Psi_2$.  This leaves 6 physical phases; therefore this model can accommodate additional CP violation beyond that present in the Standard Model.  If one also includes the down sector, and allows the two sectors to mix, there are many more physical CP-violating phases.

This CP-violation in the Higgs sector must be communicated to the matter sector for successful baryogenesis.  This can be accomplished through interactions in the bubble wall with the top quark; only the top quark Yukawa coupling is sufficiently large for the interactions to be thermal equilibrium during the phase transition.  We note that the analysis in this section is similar to \cite{McDonald:1993ey}, which considered a simpler model with two Higgs doublets and a complex singlet, of which only one doublet and the singlet acquired a nonzero vacuum expectation value.

In general, both mass eigenstates $\Psi_1$ and $\Psi_2$ couple to up-type fermions, and thus the effective Lagrangian contains terms of the form
\begin{equation}
-y_q \epsilon_{ab} T_L^a \Psi_1^{ b} \bar{t}_R - y_q^\prime \epsilon_{ab} T_L^a \Psi_2^{ b} \bar{t}_R + h.c.,
\end{equation}
where $T_L = (t_L, b_L)$ is the doublet which includes the left-handed top and bottom quarks, and $a,b$ are $SU_L(2)$ indices.  We recall that $y_q$  and $y_q^\prime$ are proportional to each other, as described by equation \eqref{eq:yukawa}, which suppresses FCNCs.   If we write the vacuum expectation values of the doublets after spontaneous symmetry breaking as
\begin{equation}
\left< \Psi_1 \right> = \begin{pmatrix}
0 \\ \xi_1 e^{\imath \theta_1}
\end{pmatrix} \qquad
\left< \Psi_2 \right> = \begin{pmatrix}
0 \\ \xi_2 e^{\imath \theta_2}
\end{pmatrix},
\end{equation}
then these terms become
\begin{equation}
\left( y_q \xi_1 e^{\imath \theta_1} + y_q^\prime \xi_2 e^{\imath \theta_2} \right) t_L \bar{t}_R + h.c.,
\end{equation}
which gives the standard top quark mass term; the potentially nonzero phase is absorbed in a rotation of the top quark.  To simplify our analysis, we will assume that $\xi_1 \gg \xi_2$; this is particularly reasonable since the mass term of $\Psi_1$ in the effective potential is negative but the mass term of $\Psi_2$ is positive.  Therefore, the dominant contribution to the top quark's mass is from the Yukawa coupling between $\Psi_1$ and the top quark, and thus we take $y_t \approx \sqrt{2} m_t \slash v = .996$.  We will define the other vacuum expectation values to be $\left< S_1 \right> = \xi_3$, $\left< S_2 \right> = \xi_4$, and $\left< V_3 \right> = 2 \left< V^0 \right> = \xi_5$.

\begin{figure}
\includegraphics[scale=.7]{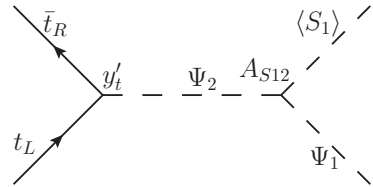}
\caption{One of the diagrams that modifies the phase of the top quark Yukawa coupling.}
\label{fig:Yukawa_Correction}
\end{figure}

When the various fields acquire nonzero vacuum expectation values, this top quark Yukawa coupling is modified, and these modifications can introduce a nonzero physical phase into the coupling.  An example of one such contribution is shown in Fig. \ref{fig:Yukawa_Correction}.  The tree-level corrections to the top-quark Yukawa coupling after spontaneous symmetry breaking are given by
\begin{align}
y_{tf} &= y_{t} + \tilde{y}_{tf} + \dfrac{y_t^\prime}{m_2^2} \left(  A_{S12} \xi_{3} + \tilde{A}_{S12} \xi_{4} + A_{V12} \xi_5 \right. \nonumber \\
& \left. + \lambda_{12}^{\prime \prime} \xi_1 \xi_2 e^{i(\theta_1 - \theta_2)} + \lambda_{12}^\prime \xi_1 \xi_2 e^{i(\theta_2 - \theta_1)} + \lambda_{S12}^\prime \xi_{3} \xi_{4}  \right. \nonumber \\
& + \lambda_{V12} \xi_5^2\left.  \right),
\end{align}
where $\tilde{y}_{tf}$ summarizes the contributions from diagrams that do not contribute a net phase.  (We have used the freedom to rotate $\Psi_1$ and $\Psi_2$ to make the parameters $\lambda_{S12}$ and $\tilde{\lambda}_{S12}$ real; we also remind the reader that we have set up our effective Lagrangian such that the singlets have zero vev before spontaneous symmetry breaking.)

Let us assume that the corrections are small with respect to $y_t$; then this corrected Yukawa coupling may be written as
\begin{align}
y_{tf} &\approx y_t e^{i \phi_f},
\end{align}
where
\begin{align}
\phi_f &= \dfrac{y_t^\prime }{y_t m_2^2} \Im \left(  A_{S12} \xi_{3} + \tilde{A}_{S12} \xi_{4} + A_{V12} \xi_5 + \lambda_{V12} \xi_5^2 \right. \nonumber \\
& \left. + \lambda_{12}^{\prime \prime} \xi_1 \xi_2 e^{i(\theta_1 - \theta_2)} + \lambda_{12}^\prime \xi_1 \xi_2 e^{i(\theta_2 - \theta_1)} + \lambda_{S12}^\prime \xi_{3} \xi_{4} \right) 
\label{eq:phase}
\end{align}
This change of phase in the Yukawa coupling can be transformed according to the standard techniques \cite{Cohen:1991iu}, \cite{Dine:1990fj}; the quarks are rotated by an amount proportional to their hypercharge to eliminate the phase, which introduces a new kinetic term for the top quark which violates CP.  The phase $\phi_t$ can be approximated as space independent, although time dependent, because the mean free path of the top quarks and gauge bosons is small compared to the scale on which $\phi_t$ varies (which is approximately the thickness of the electroweak bubble walls).  Then this additional term in the Lagrangian has the form of a chemical potential for baryon number.  Consequently, during the transition the free energy is minimized for nonzero baryon number.  If the system could evolve to the 
minimum of free energy, it would reach the baryon density
\begin{equation}
n_{B, eq} = \alpha \dfrac{T^2}{6} \dot{\phi}_t,
\end{equation}
where $\alpha$ is a constant of order 1; for a simple two-doublet model, it is $72 \slash 111$ \cite{Cohen:1991iu}.  During the phase transition, the sphaleron-induced $B+L$ violation processes drive 
the system toward this equilibrium value~\cite{Arnold:1987mh}: 
\begin{equation}
\dfrac{dn_B}{dt} = 18 \dfrac{\Gamma_{sp}}{T^3} n_{B,eq}, 
\end{equation}
but the minimum of the free energy is not reached because the transition takes place too quickly.

The sphaleron transition rate is~\cite{Kuzmin:1985mm,Dine:1990fj,Rubakov:1996vz}
\begin{equation}
\Gamma_{sp} = \begin{cases}
\kappa (\alpha_W T)^4  \quad & m_W \leq \sigma \alpha_W T \\
\gamma (\alpha_W T)^{-3} M_W^7 e^{-E_{sp} \slash T} \approx 0 \quad & m_W > \sigma \alpha_W T
\end{cases}
\end{equation}
where $\kappa$, $\sigma$, and $\gamma$ are dimensionless constants.  For the Standard Model, $\kappa$ is expected to be between .1 and 1 \cite{Ambjorn:1990pu}, while $\sigma$ is expected to be between 2 and 7 \cite{Arnold:1987mh}.  Integrating the above rate gives the baryon asymmetry produced during the phase transition
\begin{equation}
n_B = 3 \alpha \kappa \alpha_W^4 T^3 \Delta \phi_t,
\end{equation}
where $\Delta \phi_t$ is the change in the phase of the top quark Yukawa coupling during the phase transition; this is not the same as $\phi_f$  because the sphaleron $B+L$-violating interactions may go out of thermal equilibrium before the phase transition is complete.    The entropy density is
\begin{equation}
s = \dfrac{2 \pi^2}{45} g_S(T) T^3,
\end{equation}
and so baryon-to-entropy ratio after the phase transition is
\begin{equation}
\dfrac{n_B}{s} = \dfrac{135 \alpha}{2 \pi^2 g_S(T_{ew})} \kappa \alpha_W^4 \Delta \phi_t.
\end{equation}
To match the observed value of $n_B \slash s \sim 10^{-10}$, the change in phase must be of order $10^{-2}$ (assuming $g_S(T_{ew}) \sim 100$).  This is a reasonable number; given the form of equation \eqref{eq:phase}, we expect this to be satisfied for a relatively large region of parameter space.  Thus, we conclude that the electroweak phase transition in the strongly coupled MSSM can account for the observed matter asymmetry.

\section{Implications of the Triplet Vacuum Expectation Value}
\label{sec:Triplet_Problems}

We have noted in Section \ref{sec:EW_Phase_Transition} that it is an unavoidable consequence of this model that the neutral component of the hypercharge $Y=0$ Higgs triplet acquires a nonzero vacuum expectation value.  In this section, we discuss the phenomenological consequences of this, both in regards to the $\rho_0$ parameter and neutrino masses.

Models in which a single $Y=0$ Higgs triplet acquire a vacuum expectation value have been considered~\cite{Passarino:1989py,Passarino:1990nu,Lynn:1990zk,Blank:1997qa,Aoki:2011pz}; the low energy behavior of this theory was described in detail in Ref.~\cite{SekharChivukula:2007gi}.  Such models are quite constrained by precision measurements of the $\rho_0$ parameter, which is experimentally measured to be $\rho_0 = 1.0004^{+.0003}_{-.0004}$ \cite{Beringer:1900zz}.  A triplet nonzero vacuum expectation values modifies $\rho_0$ by \cite{SekharChivukula:2007gi}
\begin{equation}
\Delta \rho_0 = \dfrac{4 | \left< V^0 \right> |^2}{v_u^2}.
\end{equation}
We recall that we must have $v_u \leq 246 \; \mathrm{GeV}$; this means that we must have $|\left< V^0 \right> | = | \left< V_3 \right>| \slash 2 \lesssim 2.5 \; \mathrm{GeV}$, or equivalently,
\begin{equation}
\dfrac{|\left< V_3 \right>|}{v_u} \leq 10^{-2}.
\end{equation}
If $v_u \approx |\left< \Psi_1 \right>|$, we expect $|\left< V_3 \right>| \approx A_{V1} v_u^2 \slash m_V^2$; the above condition becomes
\begin{equation}
\dfrac{A_{V1} v_u}{m_V^2} \leq 10^{-2}.
\end{equation}
We expect $A_{V1}$ and $m_V$, like the other parameters in the effective Lagrangian, to be near the electroweak scale.  The exact values of $A_{V1}$ and $m_V$ are determined from the high energy (MSSM) Lagrangian through strong dynamics, and it is infeasible to estimate them.  It may be that a lattice calculation shows that triplet states are less strongly bound than the singlet states, and thus have larger masses, or it may be that the our model requires some fine-tuning to satisfy this condition.  We note that if $A_{V1}$ and $v_u$ are both on the 100 GeV scale, then we only require $m_V \sim \mathcal{O}(\mathrm{TeV})$.

\begin{figure}
\includegraphics[scale=.55]{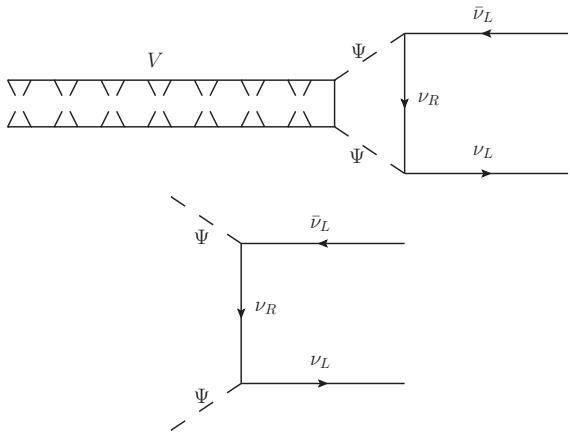}
\caption{The top diagram shows how the vector vacuum expectation value can contribute to the neutrino seesaw mass; the bottom diagram shows that this is just the closure of the standard seesaw diagram.}
\label{fig:Triplet_Neutrino}
\end{figure}

Another concern with the triplet acquiring a nonzero vacuum expectation value is that generically such vacuum expectation values may produce large neutrino masses~\cite{Mohapatra:1980yp,Schechter:1980gr,Magg:1980ut,Cheng:1980qt}.  However, our triplet, like the rest of our bound states, is comprised of squarks and true Higgs bosons, and to lowest order interacts with the neutrino only through the exchange of Higgs doublet bosons.  We observe, however, that this is just the closure of the usual seesaw mass diagram; both of these are shown in Fig. \ref{fig:Triplet_Neutrino}.  This is suppressed for the same reason the regular seesaw diagram is; the contribution is $y^2 v_u^2 \slash m_R$, which is suppressed by the large mass value of the right-handed neutrino.

Thus, although our model may require fine-tuning to satisfy current experimental constraints, the requisite fine-tuning is rather small, and the triplet vev does not generate unacceptably large neutrino masses.

\section{Collider Phenomenology}
\label{sec:collider}

Finally, we make some qualitative remarks regarding the collider phenomenology of this model.  As we have shown in Section \ref{sec:States}, this model has a rather extended Higgs sector, with numerous states.  As a result, it will be difficult to discern individual states at an experiment such as the Large Hadron Collider.    However, there may still be detectable consequences.

The gauge singlet states can be detected via deviations of the Higgs decay branching ratios from the predictions of the Standard Modal~\cite{O'Connell:2006wi,Shoemaker:2010fg}. 

Many of the Higgs states present in this model carry color charge; this is in contrast to the Standard Model and the weakly interacting MSSM, in which the Higgs sector contains only colorless states.  Again, due to the large number of such states they may be difficult to discern individually; however, these states may influence the number and structure of jets observed in high-energy scattering processes.

Secondly, we have noted in Section \ref{sec:States} the presence of a $Y=4 \slash 3$ triplet which carries color charge.  Since $SU_C(3)$ symmetry is preserved, these states must form a colorless combinations by joining with other colored particles; most frequently by pulling quarks from the quantum vacuum.  This process produces jets with integer charge.  However, some of these jets will carry charge $\pm 2$; for example, if a $\tilde{t}_L \tilde{t}_L$ bound state combines with an up quark.

Additionally, this model predicts numerous singly charged states; these arise from the extra doublet as well as the triplets.  The Standard Model, in contrast, only has an electrically neutral Higgs boson, while the MSSM has one set of $\pm 1$ charged Higgs bosons.  Furthermore, some of the singly charged states carry color charge, again in contrast to the MSSM.  Therefore, searches for charged scalar bosons may produce evidence for our model.

\section{Conclusion}

We have considered phenomenological implications of a strongly coupled realization of MSSM~\cite{Kusenko:1998yj,Giudice:1998dj,Cornwall:2012ea}.  The possibility that supersymmetry breaking could trigger electroweak phase transition leading to a low-energy effective theory with composite Higgs-like states is intriguing.  Such a strongly coupled realization of MSSM could reconcile supersymmetry with the relatively high value of the Higgs boson mass measured at LHC.  The model predicts the existence of additional Higgs bosons, which can be discovered at LHC or at a proposed future linear collider. 
The pattern of electroweak symmetry breaking is constrained so that the color $SU_C(3)$ symmetry is preserved. 

A generic prediction is the existence of a gauge-singlet Higgs boson, which gives rise to tree-level cubic terms in the effective potential.  This makes the electroweak phase transition strongly first-order for generic values of parameters.  The multi-state Higgs sector, which includes composite Higgs-like states, has a number of CP-violating phases. The combination of a first-order phase transition and new sources of CP violation opens a possibility for a successful electroweak baryogenesis. 

\section{Acknowledgments}

We thank J. M. Cornwall for very helpful, stimulating discussions. 
This work was supported by DOE Grant DE-FG03-91ER40662 and by the World Premier International Research Center Initiative (WPI Initiative), MEXT, Japan.

\appendix

\section{Effective Potential}
\label{sec:potential}

In Sections \ref{sec:EW_Phase_Transition} and \ref{sec:baryogenesis}, we made reference to the effective potential which describes our strongly interacting model before electroweak symmetry breaking.  In this appendix, we give the full potential; this is important in showing that we have sufficient freedom to set the symmetry breaking parameters as desired, and so in determining the tree-level contribution to equation \eqref{eq:phase}.  Let us call the mass eigenstates of the doublets $\Psi_1$ and $\Psi_2$, and the mass eigenstates of the singlets $S_1$ and $S_2$.  We recall that $\Psi_1$ and $\Psi_2$ are complex fields, while the others are real.  Then the potential can be written as

\begin{widetext}
\begin{align}
&V(S_1, S_2, \Psi_1, \Psi_2, V) = V_2(S_1, S_2, \Psi_1, \Psi_2, V) \nonumber \\
&\qquad + V_{3a}(\Psi_1, \Psi_2, S_1) + V_{3b}(\Psi_1, \Psi_2, S_2) + V_{3c}(S_1, S_2, V) + V_{3d}(\Psi_1, \Psi_2,V)  + V_{3e}(S_1, S_2) \nonumber \\
&\qquad + V_{4a}(\Psi_1, \Psi_2) + V_{4b} (\Psi_1, \Psi_2, S_1, S2) + V_{4c} (\Psi_1, \Psi_2, V) + V_{4d}(S_1, S_2, V) + V_{4e}(S_1, S_2),
\end{align}
where the mass terms are
\begin{equation}
V_2(S_1, S_2, \Psi_1, \Psi_2, V) = -m_1^2 \Psi_1^\dagger \Psi_1 + m_2^2 \Psi_2^\dagger \Psi_2 + m_{S1}^2 S_1^2 + m_{S2}^2 S_2^2 + m_V^2 V^T V.
\end{equation}
One of the doublet mass eigenvalues is negative due to the seesaw symmetry breaking mechanism, and we emphasize that the triplet, $V$, is real.  The cubic terms are
\begin{align}
V_{3a}(\Psi_1, \Psi_2, S_1) &= A_{S1} S_1 \Psi_1^\dagger \Psi_1 + A_{S2} S_1 \Psi_2^\dagger \Psi_2 + A_{S12} S_1 \Psi_1^\dagger \Psi_2 + h.c., \nonumber \\
V_{3b}(\Psi_1, \Psi_2, S_2) &= \tilde{A}_{S1} S_2 \Psi_1^\dagger \Psi_1 + \tilde{A}_{S2} S_2 \Psi_2^\dagger \Psi_2 + \tilde{A}_{S12} S_2 \Psi_1^\dagger \Psi_2 + h.c., \nonumber \\
V_{3c}(S_1, S_2, V) &= A_{SV} S_1 V^T V + \tilde{A}_{SV} S_2 V^T V, \nonumber \\
V_{3d}(\Psi_1, \Psi_2, V) &= A_{V1} \Psi_1^\dagger (\boldsymbol \sigma \cdot \boldsymbol V) \Psi_1 + A_{V2} \Psi_2^\dagger (\boldsymbol \sigma \cdot \boldsymbol V) \Psi_2 + A_{V12} \Psi_1^\dagger (\boldsymbol \sigma \cdot \boldsymbol V) \Psi_2 + h.c., \nonumber \\
V_{3e}(S_1, S_2) &= A_S S_1^3 + \tilde{A}_S S_2^3 + A^\prime S_1^2 S_2 + A^{\prime \prime} S_1 S_2^2.
\label{eq:Cubic_Terms}
\end{align}
Finally, the possible quartic terms are
\begin{align}
V_{4a}(\Psi_1, \Psi_2) &= \lambda_1 (\Psi_1^\dagger \Psi_1)^2 + \lambda_2 (\Psi_2^\dagger \Psi_2)^2 + \lambda_{12} (\Psi_1^\dagger \Psi_1) (\Psi_2^\dagger \Psi_2) + \lambda_{12}^\prime (\Psi_1^\dagger \Psi_2)^2 + \lambda_{12}^{\prime \prime}(\Psi_1^\dagger \Psi_2) (\Psi_2^\dagger \Psi_1) + h.c. \nonumber \\
& \quad + \lambda_1 (\Psi_1 \boldsymbol \tau \Psi_1) \cdot (\Psi_1 \boldsymbol \tau \Psi_1) + \lambda_2 (\Psi_2 \boldsymbol \tau \Psi_2) \cdot (\Psi_2 \boldsymbol \tau \Psi_2) +  \lambda_3 (\Psi_1 \boldsymbol \tau \Psi_1) \cdot (\Psi_2 \boldsymbol \tau \Psi_2) + \lambda_4 (\Psi_1 \boldsymbol \tau \Psi_2) \cdot (\Psi_1 \boldsymbol \tau \Psi_2), \nonumber \\
V_{4b}(\Psi_1, \Psi_2, S_1) &= \lambda_{S1} S_1^2 \Psi_1^\dagger \Psi_1 + \lambda_{S2} S_1^2 \Psi_2^\dagger \Psi_2 + \tilde{\lambda}_{S1} S_2^2 \Psi_1^\dagger \Psi_1 + \tilde{\lambda}_{S2} S_2^2 \Psi_2^\dagger \Psi_2 \nonumber \\
& \quad + \lambda_{S12} S_1^2 \Psi_1^\dagger \Psi_2 + \tilde{\lambda}_{S12} S_2^2 \Psi_1^\dagger \Psi_2 + \lambda_{S12}^\prime S_1 S_2 \Psi_1^\dagger \Psi_2 + h.c., \nonumber \\
V_{4c}(\Psi_1, \Psi_2, V) &= \lambda_{V1} \Psi_1^\dagger \Psi_1 V^T V + \lambda_{V2} \Psi_2^\dagger \Psi_2 V^T V + \lambda_{V12} \Psi_1^\dagger \Psi_2 V^T V +  h.c.,\nonumber \\
V_{4d}(S_1, S_2, V)&= \lambda_{SV} S_1^2 V^T V + \tilde{\lambda}_{SV} S_2^2 V^T V + \lambda_{S12} S_1 S_2 V^T V + \lambda_V (V^T V)^2, \nonumber \\
V_{4e}(S_1, S_2) &= \lambda_S S_1^4 + \tilde{\lambda}_S S_2^4 + \lambda_S^\prime S_1^3 S_2 + \tilde{\lambda}_S^\prime S_1 S_2^3 + \lambda_{SS} S_1^2 S_2^2. \nonumber \\
\label{eq:Quartic_Terms}
\end{align}
\end{widetext}
We note that the following 8 parameters are generally complex: $A_{S12}$, $\tilde{A}_{S12}$, $A_{V12}$, $\lambda_{12}^\prime$, $\lambda_{S12}$, $\tilde{\lambda}_{S12}$, $\lambda_{S12}^\prime$, and $\lambda_{V12}$.

In principle, all of the many parameters that appear in this Lagrangian are determined by the high energy theory through strong dynamics; however, it is not feasible to calculate them from first principles, although some  advanced lattice techniques could make it possible.  We expect the generic values of these parameters to lie between the supersymmetry-breaking scale and the electroweak scale.  Given the large number of parameters, we expect there to be region of parameter space in which the phase transition occurs at temperatures on the electroweak scale (which is closely related to the barrier height), and that the doublet vacuum expectation values satisfy $v_u = \sqrt{|\left< \Psi_1 \right>|^2 + |\left< \Psi_2 \right> |^2} \lesssim 246$~GeV.


\end{document}